\documentclass{WileyMSP-template}
\usepackage{cite}
\usepackage{amsmath, amssymb, physics, bm}
\usepackage{mathtools}
\usepackage{graphicx}
\usepackage{braket}
\usepackage{xcolor}
\usepackage{authblk}
\usepackage{appendix}
\usepackage{caption}
\usepackage{ulem}
\usepackage{cancel}
\usepackage{float} 
\usepackage{subcaption}  
\usepackage{physics}
\usepackage{comment}

\begin{document}

\pagestyle{fancy}
\rhead{\includegraphics[width=2.5cm]{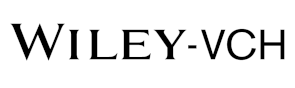}}

\title{Quantum--Plasmonic Dynamics Modeled via a Modified Langevin Noise Formalism: Numerical Studies of Single-Photon Emission and Two-Photon Interference}

\maketitle


\author{Jisang Seo}
\author{Hyunwoo Choi}
\author{Thomas E. Roth}
\author{Jie Zhu}
\author{Weng C. Chew}
\author{Dong-Yeop Na}


\dedication{}

\begin{affiliations}
Mr. Jisang Seo
Department of Electrical Engineering, Pohang University of Science and Technology, Pohang, Republic of Korea\\
Email Address: jisangseo@postech.ac.kr

Mr. Hyunwoo Choi
Department of Electrical Engineering, Pohang University of Science and Technology, Pohang, Republic of Korea\\
Email Address: hwchoi620@postech.ac.kr

Prof. Thomas E Roth
Elmore Family School of Electrical and Computer Engineering, Purdue University, West Lafayette, United States\\
Email Address: rothte@purdue.edu

Dr. Jie Zhu
Elmore Family School of Electrical and Computer Engineering, Purdue University, West Lafayette, United States\\
Email Address: zhu797@purdue.edu

Prof. Weng Cho Chew\\
Elmore Family School of Electrical and Computer Engineering, Purdue University, West Lafayette, United States\\
Email Address: wcchew@purdue.edu

Prof. Dong-Yeop Na\\
Department of Electrical Engineering, Pohang University of Science and Technology, Pohang, Republic of Korea\\
Email Address: dyna22@postech.ac.kr

\end{affiliations}



\begin{abstract}
Recent studies have established and rigorously validated a modified Langevin noise formalism that enables first-principles quantization of electromagnetic fields in open and dissipative environments~\cite{Drezet2017quantizing,Stefano2001Mode,na2023numerical}. 
Building on this foundation, a fully quantum-mechanical multimode Jaynes--Cummings framework has been developed and verified, providing an accurate description of atom--field interactions in lossy and radiative systems~\cite{choi2025non}.
In this work, we explore the potential of this formalism for nanophotonic applications by modeling representative quantum--plasmonic dynamics. 
In particular, we present detailed numerical examples for (i) two-photon interference mediated by a quantum plasmonic beam splitter, and (ii) non-Markovian dynamics of an atom located in plasmonic antennas and directional control of out-coupled single-photon fields.
These results demonstrate that the proposed modeling approach can be directly used to guide the design and optimization of plasmonic single-photon sources and beam-splitting structures. 
Moreover, the framework is broadly applicable to the analysis of linear optical components and cavity quantum electrodynamics problems in open and dissipative photonic integrated circuits.
\end{abstract}


\section{Introduction}
Practical quantum technologies---ranging from quantum communication to quantum sensing and on-chip quantum information processing---operate almost exclusively in open and dissipative electromagnetic (EM) environments. 
Unlike idealized cavity quantum electrodynamics (QED) or lossless waveguide settings, realistic photonic platforms inevitably exhibit absorption, radiation leakage, and complex modal interactions with their surrounding media. 
These effects fundamentally shape the dynamics of quantum states, influence coherence times, and determine the performance limits of integrated quantum systems.

Among various platforms, plasmonics-based quantum information science and technology (QIST) has emerged as a particularly compelling direction~\cite{tame2013quantum,Martino2014Observation,sugawara2022plasmon,heeres2013quantum,bogdanov2019overcoming}. 
Surface plasmon polaritons (SPPs) enable subwavelength confinement, ultrafast field localization, and strong enhancement of light--matter interactions, offering routes toward compact quantum circuits, ultrabright single-photon sources, and chip-scale quantum interference devices.
A representative example is the Hong--Ou--Mandel (HOM) effect, a powerful probe of photon indistinguishability in quantum optics~\cite{Hong1987measurement}. 
When two indistinguishable photons enter a 50:50 beam splitter simultaneously, the second-order correlation of the output ideally vanishes, revealing nonclassical two-photon interference. 
Recent experiments~\cite{Martino2014Observation,Dheur2016Single,Vest2017Anti} have demonstrated that indistinguishable bosonic excitations in SPP channels also exhibit HOM interference through plasmonic beam splitting, as illustrated in Fig.~\ref{fig:3D_QP_HOM}. 
Because conventional quantum-optical components (e.g., beam splitters and phase shifters) remain bulky, the preservation of bosonic interference in plasmonic media underscores their promise as a compact alternative for integrated quantum photonic platforms.
Beyond two-photon interference, plasmonic structures also enable ultrafast control, beam shaping, and directional steering of atomic spontaneous emission, as demonstrated using nanocube antennas~\cite{bogdanov2020ultrafast} and bull’s-eye antennas~\cite{hekmati2023bullseye,osorio2015k,jun2011plasmonic}. 
Such capabilities highlight the potential of plasmonic platforms to serve as highly efficient and tunable single-photon sources.

However, the same metallic and nanostructured environments that provide strong confinement also introduce substantial Ohmic and radiative losses. 
Plasmonic materials such as gold, silver, and hybrid metasurfaces possess dispersion profiles that satisfy the Kramers--Kronig relations, implying that dispersion and intrinsic Ohmic loss inevitably coexist. 
As a result, plasmonic structures become non-Hermitian: their eigenmodes lose orthonormality, their eigenfrequencies acquire imaginary parts, and ladder-operator-based diagonalization fails. 
Standard second-quantization methods for harmonic oscillators~\cite{Glauber1991quantum,scully1999quantum,CHEW2016quantum,Na2020quantum} therefore cannot be directly applied.

\begin{figure}
\centering
\includegraphics[width=0.5\linewidth]{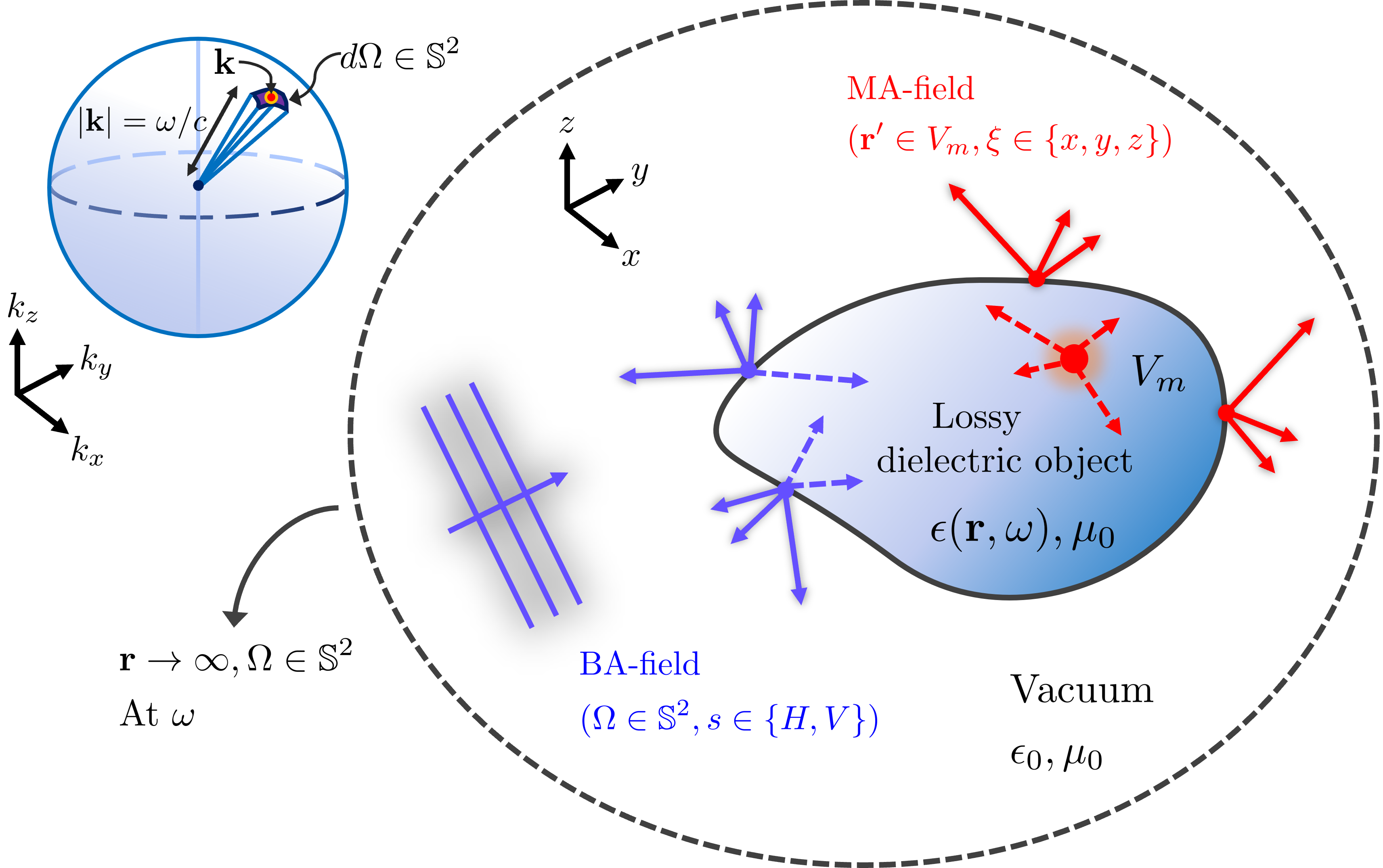}
\caption{The new Langevin noise formalism \cite{Drezet2017quantizing,Stefano2001Mode,na2023numerical} employ two different fields for the quantization of open and dissipative EM systems : (i) boundary-assisted (BA) and (ii) medium-assisted (MA) fields to balance the radiation and medium losses. Monochromatic BA and MA fields form the infinite degeneracy in terms of ($\mathbf{k}\in S_{k}$, $s\in\left\{H,V\right\}$) and ($\mathbf{r}'\in V_{m},\xi\in\left\{x,y,z\right\}$), respectively.}
\label{fig:BA--MA_schematic}
\end{figure}

To account for material absorption, the conventional Langevin noise (LN) formalism~\cite{Gruner1996Green,Dung1998three} introduces medium-assisted (MA) fields---fluctuating sources that restore correct commutation relations through the fluctuation--dissipation theorem (FDT). 
However, realistic nanophotonic systems also experience radiation loss due to open boundaries, which is not captured by MA fields alone \cite{dorier2019canonical,dorier2020critical,na2023numerical}. 
Radiation leakage constitutes an additional non-Hermitian channel and must be treated on equal footing with Ohmic dissipation.
The modified Langevin noise (M--LN) formalism~\cite{Drezet2017quantizing,Stefano2001Mode,na2023numerical,Miano2025QuantumEmitterDispersiveDielectric,Miano2025SpectralDensities} resolves this limitation by introducing boundary-assisted (BA) fields to compensate radiation loss, as illustrated in Fig.~\ref{fig:BA--MA_schematic}. 
By consistently incorporating both BA and MA contributions, the M--LN model fully satisfies the FDT in realistic open and dissipative EM environments, providing a complete and physically transparent first-principles framework for quantizing the EM field in plasmonic and nanophotonic systems.

Despite this need, many quantum-optical treatments still rely on phenomenological dissipation models or approximate mode decompositions, which cannot fully capture microscopic loss pathways arising from material absorption, radiation leakage, and the hybrid photonic continuum. 
The M--LN formalism directly overcomes these limitations by offering a rigorous and self-consistent BA--MA modal quantization, enabling accurate descriptions of atom--field and photon--field interactions in open and lossy nanophotonic environments.
Several important successes of the M--LN formalism have been demonstrated~\cite{Miano2025QuantumEmitterDispersiveDielectric,Miano2025SpectralDensities,ciattoni2024quantum,choi2025non}.  
Among them, in \cite{choi2025non}, atoms were coupled explicitly to the BA--MA continuum to investigate non-Markovian atomic dynamics in one-dimensional settings, including dissipative cavity quantum electrodynamics, superradiance and subradiance, and entanglement sudden death.  
However, extending such explicit atom--continuum coupling strategy to higher-dimensional systems is challenging, as the number of BA--MA continuum modes grows rapidly and becomes computationally intractable compared to the manageable sampling required in the one-dimensional case.  
To address this obstacle, recent work introduced a computationally efficient approach for treating non-Markovian spontaneous emission dynamics in open~\cite{zhou2024simulating} and dissipative EM environments~\cite{choi2025atom}.  
As shown in~\cite{choi2025atom}, this method relies on a memory-kernel formulation directly linked to the imaginary part of the dyadic Green’s function, enabling seamless integration with computational electromagnetic (CEM) solvers such as the finite-element method and the auxiliary differential equation finite-difference time-domain method with total-field/scattered-field implementation. 
These studies primarily focused on simple open and dissipative cavity-QED scenarios.

In this work, we leverage the M--LN formalism to model practical and technologically relevant scenarios, focusing on two central quantum--plasmonic phenomena.

First, within the Heisenberg picture, we analyze two-photon interference in a plasmonic beam-splitting platform.  
For a plasmonic structure embedded in free space, the electric-field operator is expanded in terms of BA--MA field modes.  
We model two initially separable single photons incident from free space onto a grating coupler.  
Accordingly, each incident single photon is represented as a single-quantum excitation expressed through a properly weighted linear superposition of BA field modes, and the total two-photon state is given by the tensor product of the two individual single-photon states.  
Under the assumption of zero temperature, thermal excitations in the plasmonic structure are negligible, and the initial MA field modes contain no quanta.  
Since BA and MA fields do not interact, only the BA channels contribute in this setting.  
By evaluating the second-order correlations between out-coupled single-photon fields and surface-plasmon--polariton modes, we demonstrate Hong--Ou--Mandel--type interference.  
Notably, we show that the resulting expressions for the second-order correlation and the single-photon detection probability can be computed directly from two time-domain classical EM simulations corresponding to the two incident single-photon wavepackets.
Second, within the Schr\"{o}dinger picture, we investigate single-photon spontaneous emission in a plasmonic grating-antenna environment.  
By combining M--LN quantization with a multimode Jaynes--Cummings (MMJC) model, we derive a modified MMJC formulation that remains computationally efficient while fully capturing atom--plasmon--radiation coupling.  
We compute both atomic population dynamics and out-coupled single-photon probability amplitudes, revealing how plasmonic-antenna geometries enable beam shaping and directional steering of the emitted photon.  
In particular, we propose an efficient method that explicitly incorporates BA--MA mode coupling to the two-level system (TLS) while avoiding the prohibitive computational cost associated with sampling the full continuum degeneracy space of BA--MA modes.  
The key idea is to lump out the continuum degeneracy degrees of freedom (DoFs) and define effective DoFs that depend only on frequency, which are then coupled to the TLS.  
This eliminates the need for sampling the large BA--MA mode space and yields a highly efficient computational framework.  
Using this approach, we numerically demonstrate how a TLS embedded inside a slit exhibits distinct population dynamics and directional emission depending on whether the surrounding plasmonic grating has grooves or not, and depending on whether the geometry is symmetric or asymmetric.

Together, these examples illustrate that the M--LN formalism offers a fully first-principles, versatile, and numerically efficient route for modeling quantum optical phenomena in realistic plasmonic systems.  

Note that all simulations in this paper were performed using natural units where $\epsilon_0 = \mu_0 = c = \hbar = 1$.

\section{Modified Langevin Noise Formalism}
\subsection{Theoretical Background}
Consider a dispersive and lossy dielectric object occupying a volume $V_m$ in free space, whose permittivity is given by
\begin{flalign}
\epsilon(\mathbf{r},\omega)
=
\epsilon_0 \epsilon_r(\mathbf{r},\omega)
=
\epsilon_0 \bigl[1+\chi_e(\mathbf{r},\omega)\bigr],
\end{flalign}
where $\chi_e(\mathbf{r},\omega)$ is the electric susceptibility satisfying the Kramers--Kr\"onig relations required for causality.
Within the M--LN formalism, the EM field in such an open and dissipative environment is represented by a complete set of boundary-assisted (BA) and medium-assisted (MA) fields.  
The positive-frequency part of the electric field operator is expanded as
\begin{flalign}
\hat{\mathbf{E}}^{(+)}(\mathbf{r},t)
=
\hat{\mathbf{E}}^{(+)}_{\text{(BA)}}(\mathbf{r},t)
+
\hat{\mathbf{E}}^{(+)}_{\text{(MA)}}(\mathbf{r},t),
\end{flalign}
where the BA-field contribution takes the form
\begin{flalign}
\hat{\mathbf{E}}^{(+)}_{\text{(BA)}}(\mathbf{r},t)
=
i
\int_{0}^{\infty} d\omega
\int_{\mathbb{S}^{2}} k^{2} d\Omega 
\sum_{s \in \{H,V\}}
\mathbf{E}_{\text{B}}(\mathbf{r};\Omega,s,\omega)
\,
\hat{a}_{\text{B}}(\Omega,s,\omega)
e^{-i\omega t},
\end{flalign}
and the MA-field contribution is expressed as
\begin{flalign}
\hat{\mathbf{E}}^{(+)}_{\text{(MA)}}(\mathbf{r},t)
=
i
\int_{0}^{\infty} d\omega
\int_{V_m} d\mathbf{r}'
\sum_{\xi \in \{x,y,z\}}
\mathbf{E}_{\text{M}}(\mathbf{r};\mathbf{r}',\xi,\omega)
\,
\hat{a}_{\text{M}}(\mathbf{r}',\xi,\omega)
e^{-i\omega t}.
\end{flalign}
Here,  
$\hat{a}_{\text{B}}$ and $\hat{a}_{\text{M}}$ denote annihilation operators associated with BA and MA fields, respectively;  
$\Omega$ is the solid angle on the unit sphere $\mathbb{S}^2$;  
$s$ denotes the plane-wave polarization;  
$\xi$ represents the dipole orientation within the lossy medium;  
and $\mathbf{E}_{\text{B}}$ and $\mathbf{E}_{\text{M}}$ correspond to the BA and MA field mode functions.
The mode functions entering the expansion are given by
\begin{flalign}
\mathbf{E}_{\text{B}}(\mathbf{r};\Omega,s,\omega)
&=
\frac{1}{(2\pi)^{d/2}}
\boldsymbol{\Phi}_{\text{tot}}(\mathbf{r};\Omega,s,\omega)
\sqrt{\frac{\hbar \omega}{2}}, \\
\mathbf{E}_{\text{M}}(\mathbf{r};\mathbf{r}',\xi,\omega)
&=
k^{2}
\overline{\mathbf{G}}_{E}(\mathbf{r};\mathbf{r}',\omega)
\cdot \hat{\xi}
\sqrt{
\frac{
\hbar\, \mathrm{Im}\!\left[\epsilon(\mathbf{r}',\omega)\right]
}{
\pi \epsilon_0}
},
\end{flalign}
where $d$ denotes dimension, and $k=\omega/c$ is the free-space wavenumber,  
$\boldsymbol{\Phi}_{\text{tot}}$ is the total field resulting from the incident plane wave and its scattered field,  
and $\overline{\mathbf{G}}_{E}$ denotes the dyadic Green's function for the electric field.

The BA--MA field modes possess the following important properties:
\begin{itemize}
\item BA fields originate from radiation fluctuations in free space. Each continuum degeneracy corresponds to an incident plane wave arriving from the infinitely extended boundary. Thus, a single degenerate BA mode is composed of an incident plane wave together with the associated scattered fields generated by dielectric objects. Consequently, BA fields exhibit degeneracy with respect to the plane-wave propagation direction (solid angle $\Omega$ in $k$-space) and polarization $s$.

\item MA fields arise from fluctuations associated with medium Ohmic loss. Each continuum degeneracy is determined by the position and orientation of a noise point-current source residing inside the absorbing dielectric region $V_m$. Accordingly, a single degenerate MA mode corresponds to the EM field generated by such a point-current source within the medium. Therefore, MA fields exhibit degeneracy with respect to the source location ($\mathbf{r}'\in V_m$) and orientation $\xi$.

\item BA and MA field modes do not interact with one another; their fluctuation channels are independent and mutually decoupled.

\item The combined set of BA--MA field modes satisfies transverse modal completeness \cite{choi2025atom}, providing a complete basis for expanding the electric field operator in open and dissipative EM environments, i.e., 
\begin{flalign}
\frac{\omega^2}{c \hbar}
\text{Im}\left[\overline{\mathbf{G}}(\mathbf{r};\mathbf{r}',\omega)\right]
=
\sum_{s \in \{H,V\}}
\int_{\mathbb{S}^2} d\Omega~\mathbf{E}(\mathbf{r};\Omega,s,\omega) \otimes \mathbf{E}^{*}(\mathbf{r}';\Omega,s,\omega)
+
\sum_{\zeta=\left\{x,y,z\right\}}\int_{V_{m}}d\mathbf{r''}
\mathbf{E}(\mathbf{r};\mathbf{r}'',\zeta,\omega) \otimes \mathbf{E}^{*}(\mathbf{r}';\mathbf{r}'',\zeta,\omega)
\label{eqn:TMC_BAMA}
\end{flalign}
\end{itemize}

More specifically, each BA field mode contains the total field consisting of
\begin{flalign}
\boldsymbol{\Phi}_{\text{tot}}(\mathbf{r};\Omega,s,\omega)
=
\boldsymbol{\Phi}_{\text{inc}}(\mathbf{r};\Omega,s,\omega)
+
\boldsymbol{\Phi}_{\text{sca}}(\mathbf{r};\Omega,s,\omega)
\end{flalign}
where $\boldsymbol{\Phi}_{\text{inc}}(\mathbf{r};\Omega,s,\omega) = \hat{\mathbf{e}}_{s} e^{i\mathbf{k}\cdot\mathbf{r}}$ and the resulting scattered field can be found from a plane-wave-scattering problem stated by
\begin{flalign}
\nabla\times\nabla\times
\boldsymbol{\Phi}_{\text{sca}}(\mathbf{r};\Omega,s,\omega)
-
\frac{\omega^2}{c^2}\epsilon_r(\mathbf{r},\omega)\boldsymbol{\Phi}_{\text{sca}}(\mathbf{r};\Omega,s,\omega)
=
\frac{\omega^2}{c^2}
\chi_e(\mathbf{r},\omega)
\boldsymbol{\Phi}_{\text{inc}}(\mathbf{r};\Omega,s,\omega)
\label{eqn:plane-wave-scattering}
\end{flalign}

On the other hand, each MA field mode corresponds to an EM fields created by a point current source in dispersive and absorbing dielectric objects; hence, it has a dyadic Green's function governed by
\begin{flalign}
\nabla\times\nabla\times
\overline{\mathbf{G}}_E(\mathbf{r};\mathbf{r}',\omega)
-
\frac{\omega^2}{c^2}
\epsilon_r(\mathbf{r},\omega)
\overline{\mathbf{G}}_E(\mathbf{r};\mathbf{r}',\omega)
=
\overline{\mathbf{I}}\delta(\mathbf{r}-\mathbf{r}').
\label{eqn:DGF}
\end{flalign}

Two different ladder operators, i.e., $\hat{a}_{\text{B}}(\Omega,s,\omega)$ and $\hat{a}_{\text{M}}(\mathbf{r}',\xi,\omega)$, are associated with BA and MA fields, respectively, satisfy the standard bosonic commutator relations, 
\begin{flalign}
\left[\hat{a}^{\dag}_{\text{B}}(\Omega,s,\omega),\hat{a}_{\text{B}}(\Omega',s',\omega')\right]
&=
\hat{I}\delta(\Omega-\Omega')\delta_{s,s'}\delta(\omega-\omega'),
\\
\left[\hat{a}^{\dag}_{\text{M}}(\mathbf{r}',\xi,\omega),\hat{a}_{\text{M}}(\mathbf{r}_m,\xi',\omega')\right]
&=
\hat{I}\delta(\mathbf{r}'-\mathbf{r}_m)\delta_{\xi,\xi'}\delta(\omega-\omega'),
\\
\left[\hat{a}^{\dag}_{\text{M}}(\mathbf{r}',\xi,\omega),\hat{a}_{\text{B}}(\Omega,s,\omega)\right] &= 0
\end{flalign}
and take multimode Fock states as eigenstates of their number operators.

The resulting field Hamiltonian operator can be written by the two diagonalizing ladder operators \cite[Eqn. (4.37)]{Drezet2017quantizing}
\begin{flalign}
\hat{H}
&=
\int_{0}^{\infty}d\omega
\int_{\mathbb{S}^{2}} k^2 d\Omega
\sum_{s \in \{H,V\}}
\hbar\omega
\hat{a}_{\text{B}}^{\dag}(\Omega,s,\omega)
\hat{a}_{\text{B}}(\Omega,s,\omega)
+
\int_{0}^{\infty}d\omega 
\int_{V_{m}}d\mathbf{r}'
\sum_{\xi\in\left\{x,y,z\right\}}
\hbar\omega
\hat{a}_{\text{M}}^{\dag}(\mathbf{r}',\xi,\omega)
\hat{a}_{\text{M}}(\mathbf{r}',\xi,\omega).
\label{eqn:Ham}
\end{flalign}

Replacing all the continuum degeneracy indices (i.e., $\Omega$, $s$, $\mathbf{r}'$, and $\xi$) with a single continuum degeneracy index $\lambda$, one can write the unified expression by
\begin{flalign}
\hat{\mathbf{E}}^{(+)}(\mathbf{r},t)
=
i \int_{0}^{\infty} d\omega \int_{\mathcal{D}_{\lambda}} \mathbf{E}_{\omega,\lambda}(\mathbf{r})\hat{a}_{\omega,\lambda} e^{-i\omega t}
\end{flalign}
with the field Hamiltonian
\begin{flalign}
\hat{H}_{\text{field}} = \int_{0}^{\infty} d\omega \int_{\mathcal{D}_{\lambda}}\hbar\omega \hat{a}^{\dag}_{\omega,\lambda}\hat{a}_{\omega,\lambda}.
\end{flalign}

\section{Two-Photon Interference}
We consider a two-dimensional simulation scenario in which two independent single photons impinge on a plasmonic structure from free space. Each photon is assumed to be both spatially and temporally localized. A natural way to model such a single-photon excitation is to represent its quantum state as a superposition of single-photon Fock states distributed over multiple EM field modes, with the corresponding probability amplitudes determining the weight of each mode. These probability amplitudes play a role analogous to the spectral envelope of a classical wavepacket and encode the spatio-temporal localization characteristic of the single photon.
In this work, we model each single photon using a Gaussian wavepacket, which provides a convenient and physically well-motivated form for specifying the modal distribution of the excitation.
It is important to emphasize that, since the incident single photons originate from free space, their mode expansion must be expressed exclusively in terms of BA field modes. 
Assuming a background temperature of 0~[K], we neglect any initial excitations associated with MA field modes, which would otherwise appear as thermal (mixed) states at finite temperature.

Note that in our two-dimensional simulations, we consider only the $\mathrm{TE}^{z}$ (or $V$) polarization.
Because the analysis is restricted to a single polarization, the polarization DoF $s$ in the BA field modes is omitted in the following discussion.
In addition, in two-dimensional space, the solid-angle variable $\Omega$ reduces to the azimuthal angle $\phi$.
Accordingly, the differential measure appearing in the BA field mode expansion must be replaced by the two-dimensional $\mathbf{k}$-space measure
\begin{flalign}
d\mathbf{k}
= k\, dk\, d\phi
= \frac{\omega}{c^2}\, d\omega\, d\phi,
\end{flalign}
where we used the relation $\omega = c k$ and $d\Omega \equiv d\phi$ in two dimensions.

\subsection{Initial quantum state}
An initial quantum state for a single photon riding on a wavepacket (i.e., spatially and temporally localized) coming from infinity can be modeled as
\begin{flalign}
\ket{\Psi^{(1)}_{t_0}}
=
\int_{0}^{\infty} d\omega
\int_{0}^{2\pi}
\frac{\omega}{c^2}\, d\phi\,
\tilde{g}(\phi,\omega)\,
\ket{1_{\mathrm{B}}(\phi,\omega)}
=
\int_{0}^{\infty} d\omega
\int_{0}^{2\pi}
\frac{\omega}{c^2}\, d\phi\,
\tilde{g}(\phi,\omega)\,
\hat{a}^{\dagger}_{\mathrm{B}}(\phi,\omega)
\ket{0}.
\end{flalign}

where $\tilde{g}$ denotes the spectral probability amplitudes associated with the single-photon Fock state $\ket{1_{\text{B}}}$ of the BA field modes.
Due to the probability normalization condition, the spectral amplitude must satisfy
\begin{flalign}
\int_{0}^{\infty} d\omega
\int_{0}^{2\pi}
\frac{\omega}{c^2}\, d\phi\,
\left|\tilde{g}(\phi,\omega)\right|^2 = 1.
\end{flalign}

Each BA field mode inherently consists of an incoming plane wave---specified by a particular propagation direction and polarization---and the corresponding scattered component.
Therefore, an appropriate linear superposition of BA field modes naturally represents the time-evolving wavepacket that originates from free space, impinges on a target, and undergoes scattering.
Consequently, a spatially and temporally localized incident single photon should be modeled as a linear superposition of single-quanta Fock states associated with the BA field modes.

When the initial Gaussian localization function, which specifies the spatial profile of the single photon in two-dimensional space, is given by
\begin{flalign}
g_0(\mathbf{r})
=
\exp\!\left(-\frac{|\mathbf{r}-\mathbf{r}_0|^2}{2\sigma_0^2}\right),
\end{flalign}
where $\mathbf{r}_0$ denotes the center of the localization function and $\sigma_0$ characterizes its spatial width.
The corresponding spectral component is obtained through
\begin{flalign}
\tilde{g}_0(\mathbf{k})
=
\int_{\mathbb{R}^2} 
d\mathbf{r}
g_0(\mathbf{r})\,
\exp\!\left(-i\mathbf{k}\cdot\mathbf{r}\right)
\end{flalign}
where, in the two-dimensional space $\mathbb{R}^2$, a wavevector $\mathbf{k} = k \left(\cos\phi\hat{\mathbf{x}} + \sin\phi\hat{\mathbf{y}}\right)$.
Using this localization spectrum, the spectral probability amplitude of a Gaussian wavepacket centered at $\mathbf{k}_p$ can be written as
\begin{flalign}
\tilde{g}(\mathbf{k})
=
\tilde{g}_0(\mathbf{k})\,
\exp\!\left(-\frac{|\mathbf{k}-\mathbf{k}_p|^2}{2\tilde{\sigma}_0^2}\right),
\end{flalign}
where $\mathbf{k}_p$ specifies the central propagation direction of the Gaussian wavepacket, and $\tilde{\sigma}_0$ characterizes its spectral width.

When the above spectral amplitude is taken in the plane-wave basis (which corresponds to the BA field modes in the absence of scattering targets), the resulting quantum state models the free-space propagation of the initial Gaussian wavepacket. 
In contrast, when the same spectral amplitude is expanded in the full BA field modes in the presence of scattering objects, the time evolution naturally describes a wavepacket incident from free space, interacting with the targets, and subsequently scattering back into the environment. 
Therefore, in order to model the plasmonic Hong--Ou--Mandel effect---where two separable single photons, each riding on a Gaussian wavepacket, propagate from free space toward the plasmonic platform---we employ this strategy to construct the initial quantum states of the incident photons.

With this prescription for generating a single-photon state, we can construct a two-photon incident state. 
If the two photons are unentangled, the initial state must be separable and can therefore be written as the tensor product of the two single-photon states:
\begin{flalign}
\ket{\Psi^{(2)}(t_0)}
&=
\ket{\Psi^{(1\!-\!L)}(t_0)} \otimes \ket{\Psi^{(1\!-\!R)}(t_0)}
\nonumber\\
&=
\left[
\int_{0}^{\infty} d\omega
\int_{0}^{2\pi} \frac{\omega}{c^2}\, d\phi\;
\tilde{g}_{(\mathrm{L})}(\phi,\omega)
\ket{1_{\mathrm{B}}(\phi,\omega)}
\right]
\otimes
\left[
\int_{0}^{\infty} d\omega
\int_{0}^{2\pi} \frac{\omega}{c^2}\, d\phi\;
\tilde{g}_{(\mathrm{R})}(\phi,\omega)
\ket{1_{\mathrm{B}}(\phi,\omega)}
\right],
\end{flalign}
where the subscripts L and R label the left- and right-incident photons, respectively.

\subsection{Second-order correlation and single-photon detection probability}

We evaluate the second-order correlation function
\[
g^{(2)}(\mathbf{r}_\alpha,t_\alpha;\mathbf{r}_\beta,t_\beta)
=
\frac{\Braket{A|A}}{B_\alpha B_\beta},
\]
where
\begin{flalign}
\ket{A}
&=
\hat{\mathbf{E}}^{(+)}_{\mathrm{(BA)}}(\mathbf{r}_\beta,t_\beta)
\cdot
\hat{\mathbf{E}}^{(+)}_{\mathrm{(BA)}}(\mathbf{r}_\alpha,t_\alpha)
\ket{\Psi^{(2)}_{t_0}},
\\[4pt]
B_\zeta
&=
\mel{\Psi^{(2)}_{t_0}}
{
\hat{\mathbf{E}}^{(+)}_{\mathrm{(BA)}}(\mathbf{r}_\zeta,t_\zeta)
\cdot
\hat{\mathbf{E}}^{(+)}_{\mathrm{(BA)}}(\mathbf{r}_\zeta,t_\zeta)
}
{\Psi^{(2)}_{t_0}},
\end{flalign}
for $\zeta=\alpha$, $\beta$.
Using the orthonormality of Fock states and bosonic commutation relations,
$\ket{A}$ becomes
\begin{flalign}
\ket{A}
&=
\left[
\int_{0}^{\infty} d\omega
\int_{0}^{2\pi} \frac{\omega}{c^2} d\phi\;
\tilde{g}_{(\mathrm{L})}(\phi,\omega)\,
\mathbf{E}_{\mathrm{B}}(\mathbf{r}_\beta;\phi,\omega)\,
e^{-i\omega t_\beta}
\right]
\cdot
\left[
\int_{0}^{\infty} d\omega
\int_{0}^{2\pi} \frac{\omega}{c^2} d\phi\;
\tilde{g}_{(\mathrm{R})}(\phi,\omega)\,
\mathbf{E}_{\mathrm{B}}(\mathbf{r}_\alpha;\phi,\omega)\,
e^{-i\omega t_\alpha}
\right]
\ket{0}
\nonumber\\[4pt]
&+
\left[
\int_{0}^{\infty} d\omega
\int_{0}^{2\pi} \frac{\omega}{c^2} d\phi\;
\tilde{g}_{(\mathrm{R})}(\phi,\omega)\,
\mathbf{E}_{\mathrm{B}}(\mathbf{r}_\beta;\phi,\omega)\,
e^{-i\omega t_\beta}
\right]
\cdot
\left[
\int_{0}^{\infty} d\omega
\int_{0}^{2\pi} \frac{\omega}{c^2} d\phi\;
\tilde{g}_{(\mathrm{L})}(\phi,\omega)\,
\mathbf{E}_{\mathrm{B}}(\mathbf{r}_\alpha;\phi,\omega)\,
e^{-i\omega t_\alpha}
\right]
\ket{0}.
\end{flalign}
Likewise, the single-photon detection probability becomes
\begin{flalign}
B_\zeta
&=
\left|
\int_{0}^{\infty} d\omega
\int_{0}^{2\pi} \frac{\omega}{c^2} d\phi\;
\tilde{g}_{(\mathrm{L})}(\phi,\omega)\,
\mathbf{E}_{\mathrm{B}}(\mathbf{r}_\zeta;\phi,\omega)\,
e^{-i\omega t_\zeta}
\right|^2
+
\left|
\int_{0}^{\infty} d\omega
\int_{0}^{2\pi} \frac{\omega}{c^2} d\phi\;
\tilde{g}_{(\mathrm{R})}(\phi,\omega)\,
\mathbf{E}_{\mathrm{B}}(\mathbf{r}_\zeta;\phi,\omega)\,
e^{-i\omega t_\zeta}
\right|^2
\nonumber\\
&\quad+
\int_{0}^{\infty} d\omega
\int_{0}^{2\pi} \frac{\omega}{c^2} d\phi\;
|\tilde{g}_{(\mathrm{L})}(\phi,\omega)|^2
+
\int_{0}^{\infty} d\omega
\int_{0}^{2\pi} \frac{\omega}{c^2} d\phi\;
|\tilde{g}_{(\mathrm{R})}(\phi,\omega)|^2 .
\end{flalign}
Since each single-photon wavepacket is normalized,
\[
\int_{0}^{\infty} d\omega
\int_{0}^{2\pi} \frac{\omega}{c^2} d\phi\;
|\tilde{g}_{(\xi)}(\phi,\omega)|^2
=
1,
\]
for $\xi=$ L, R.
And, $B_\zeta$ reduces to
\begin{flalign}
B_\zeta
&=
\left|
\int_{0}^{\infty} d\omega
\int_{0}^{2\pi} \frac{\omega}{c^2} d\phi\;
\tilde{g}_{(\mathrm{L})}(\phi,\omega)\,
\mathbf{E}_{\mathrm{B}}(\mathbf{r}_\zeta;\phi,\omega)\,
e^{-i\omega t_\zeta}
\right|^2
+
\left|
\int_{0}^{\infty} d\omega
\int_{0}^{2\pi} \frac{\omega}{c^2} d\phi\;
\tilde{g}_{(\mathrm{R})}(\phi,\omega)\,
\mathbf{E}_{\mathrm{B}}(\mathbf{r}_\zeta;\phi,\omega)\,
e^{-i\omega t_\zeta}
\right|^2
+
2,
\end{flalign}
for $\zeta=\alpha,\beta$.

The physical meaning of $\braket{A|A}$ is the probability of coincident photodetections 
at $(\mathbf{r}_\alpha,t_{\alpha})$ and $(\mathbf{r}_\beta,t_{\beta})$, 
while $B_\alpha$ (or $B_\beta$) describes the probability of detecting a single photon 
at $(\mathbf{r}_\alpha,t_{\alpha})$ (or $(\mathbf{r}_\beta,t_{\beta})$), respectively.
It is interesting to observe that the coincidence probability in the numerator can be computed purely from a classical EM viewpoint, 
since it is given by the sum of products of two independently propagated wavepacket envelopes.
Likewise, the denominator, describing single-photon detection probabilities, can also be evaluated classically 
from the intensities of the two wavepacket evolutions, together with the integrated spectral probabilities.
Therefore, the second-order correlation can be computed from two independent classical EM simulations of the two incident wavepackets. 
If we denote the classically propagated field envelope as
\begin{flalign}
\mathbf{E}^{(+)}(\mathbf{r}_\zeta,t_\zeta;\xi)
\triangleq
\int_{0}^{\infty} d\omega
\int_{0}^{2\pi} \frac{\omega}{c^2} d\phi\;
\tilde{g}_{(\xi)}(\phi,\omega)\,
\mathbf{E}_{\mathrm{B}}(\mathbf{r}_\zeta;\phi,\omega)\,
e^{-i\omega t_\zeta},
\end{flalign}
then the second-order correlation function is written compactly as
\begin{flalign}
g^{(2)}(\mathbf{r}_\alpha,t_\alpha;\mathbf{r}_\beta,t_\beta)
=
\frac{
\left|
\mathbf{E}^{(+)}(\mathbf{r}_\beta,t_\beta;\mathrm{L}) \cdot
\mathbf{E}^{(+)}(\mathbf{r}_\alpha,t_\alpha;\mathrm{R})
+
\mathbf{E}^{(+)}(\mathbf{r}_\beta,t_\beta;\mathrm{R}) \cdot
\mathbf{E}^{(+)}(\mathbf{r}_\alpha,t_\alpha;\mathrm{L})
\right|^2
}{
\Big(
\left|\mathbf{E}^{(+)}(\mathbf{r}_\alpha,t_\alpha;\mathrm{L})\right|^2
+
\left|\mathbf{E}^{(+)}(\mathbf{r}_\alpha,t_\alpha;\mathrm{R})\right|^2
+
2
\Big)
\Big(
\left|\mathbf{E}^{(+)}(\mathbf{r}_\beta,t_\beta;\mathrm{L})\right|^2
+
\left|\mathbf{E}^{(+)}(\mathbf{r}_\beta,t_\beta;\mathrm{R})\right|^2
+
2
\Big)
}.
\end{flalign}
The out-coupled single-photon detection probability is then given by
\begin{flalign}
\mathrm{SPDP}(\mathbf{r},t)
=
\left|\mathbf{E}^{(+)}(\mathbf{r},t;\mathrm{L})\right|^2
+
\left|\mathbf{E}^{(+)}(\mathbf{r},t;\mathrm{R})\right|^2
+
2.
\end{flalign}

\subsection{Simulation results of plasmonic Hong--Ou--Mandel effects}

\begin{figure}[t]
\centering
\includegraphics[width=.5\linewidth]{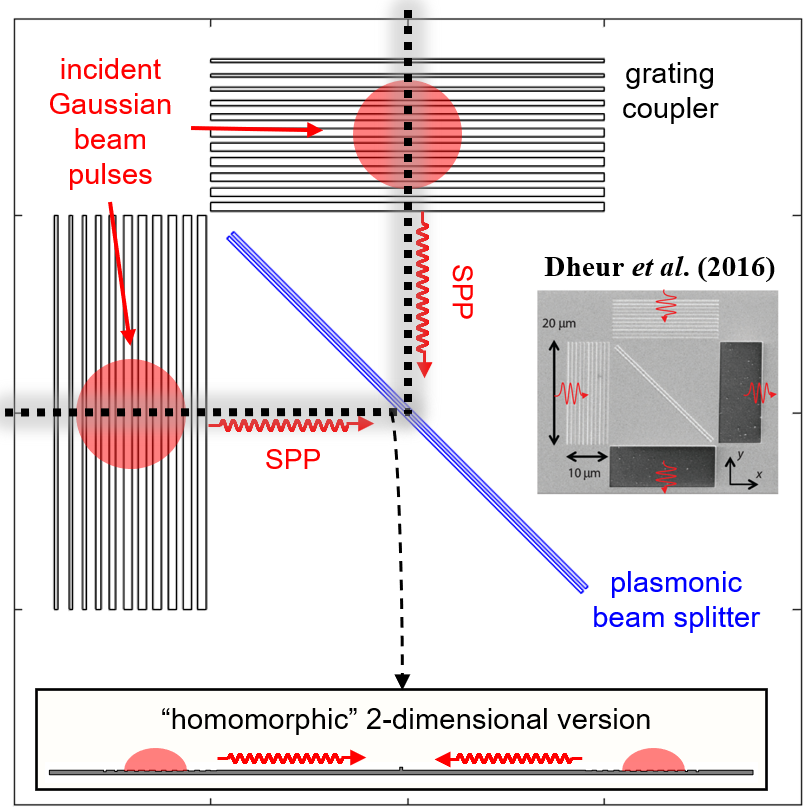}
\caption{A plasmonic platform for observing quantum plasmonic Hong--Ou--Mandel effects. 
The inset SEM image (right) of the photon-to-surface-plasmon-polariton launcher was reproduced from \cite{Dheur2016Single}. 
The black dashed contour indicates the ``homomorphic'' two-dimensional platform used in our simulations.}
\label{fig:3D_QP_HOM}
\end{figure}

\begin{figure}[t]
\centering
\includegraphics[width=0.75\linewidth]{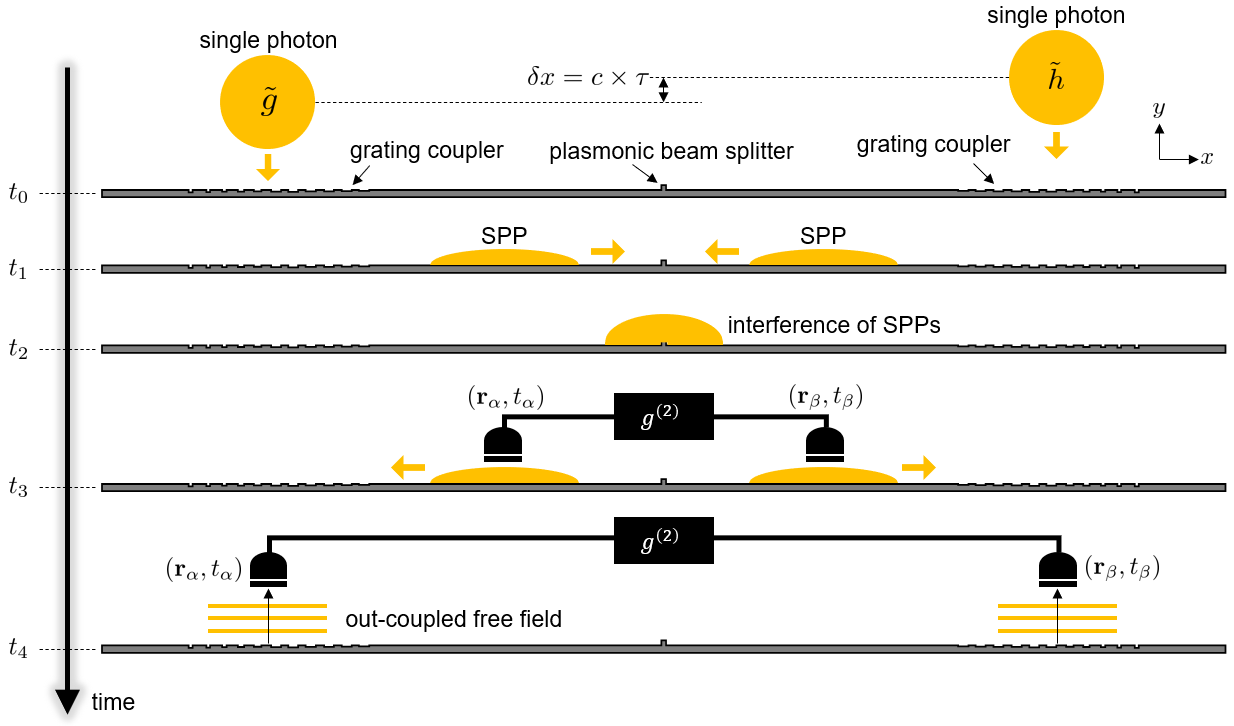}
\caption{Numerical simulation scenario for observing the quantum plasmonic Hong--Ou--Mandel effect in two dimensions.}
\label{fig:Scenario}
\end{figure}

Based on the developed numerical framework, we perform two-dimensional simulations to observe 
plasmonic Hong--Ou--Mandel (HOM) interference---a model that is homomorphic to the full three-dimensional
configuration \cite{Dheur2016Single} shown in Fig.~\ref{fig:3D_QP_HOM}.
The two-dimensional simulation scenario is illustrated in Fig.~\ref{fig:Scenario}. 
The plasmonic platform consists of a gold (Au) thin film containing (i) two grating couplers on the left and right 
and (ii) a single-ridge plasmonic beam splitter (BS) located at the center.  
The grating design parameters follow \cite{Baron2011Compact}, 
and the ridge width and height are chosen to be $200\,\mathrm{nm}$, giving approximately $50{:}50$ splitting.  
The dispersive permittivity $\epsilon(\mathbf{r},\omega)$ of Au is modeled using the experimental data in \cite{McPeak2015}.

We consider two unentangled single photons in vacuum incident on the grating couplers. 
Each photon has a center wavelength of $800\,\mathrm{nm}$ and an $x$-polarized electric field that efficiently excites surface plasmon polaritons (SPPs).  
We assume that each single photon has a temporal full width at half maximum (FWHM) of approximately $0.019~\mathrm{ps}$.
The excited SPPs propagate toward the plasmonic BS, undergo plasmonic interference, and are subsequently converted back into out-coupled photons by the grating couplers.  
The out-coupled photons are then detected at two photodetectors (at time $t_4$ in Fig.~\ref{fig:Scenario}), 
thereby generating a second-order correlation signal.  
In addition, we evaluate second-order correlations directly from the SPP fields before their conversion into free-space photons (at time $t_3$ in Fig.~\ref{fig:Scenario}) to verify that SPPs, being bosonic quasiparticles, can also exhibit HOM visibility.

In order to analyze the plasmonic HOM visibility, and following the numerical framework developed in the previous section, we perform two independent time-evolving Gaussian wavepacket simulations. 
By shifting the initial position of the right-incident wavepacket, whose spectral amplitude is denoted by $\tilde{h}$, we introduce a controllable time delay between the two single-photon inputs.  
More specifically, for each frequency component within the bandwidth of the single-photon wavepacket (corresponding to its temporal FWHM), we run a frequency-domain finite-element method (FEM) simulation of a Gaussian beam centered at the initial photon location.  
The full time evolution of the Gaussian wavepacket is then obtained by weighting these FEM solutions with a Gaussian spectral distribution.  
Using this procedure, we retrieve the positive-frequency fields $\mathbf{E}^{(+)}(\mathbf{r},t;\xi)$ for $\xi=\mathrm{L},\mathrm{R}$.

\begin{figure}[t]
\centering
\includegraphics[width=.5\linewidth]{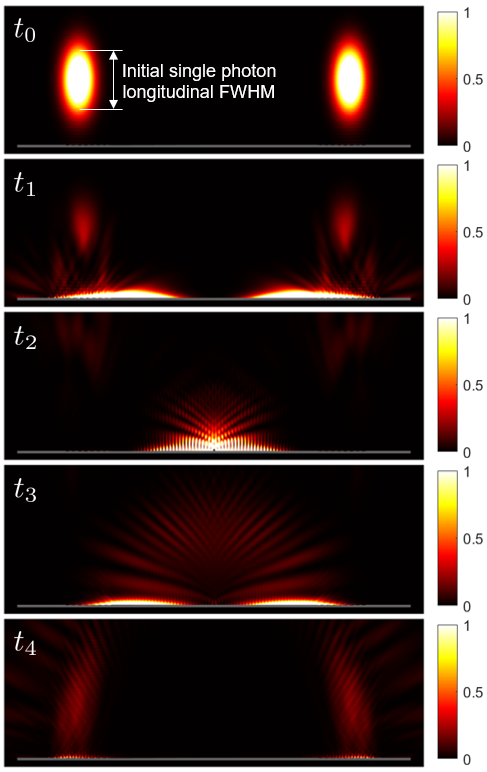}
\caption{Single-photon detection probability at successive time instants 
$t=t_0,t_1,t_2,t_3,t_4$ in the presence of the plasmonic beam splitter.  
The sequence shows photon-to-SPP conversion, SPP propagation and interference, and reconversion into out-coupled photons.}
\label{fig:Time_Evolution_E_field}
\end{figure}
Fig.~\ref{fig:Time_Evolution_E_field} shows the evolution of the single-photon detection probability in the presence of the BS.  
The two localized input photons are first converted into SPPs via the grating couplers, propagate toward the BS where plasmonic interference occurs,  
and then reconvert into out-coupled photons.

\begin{figure}[t]
\centering
\includegraphics[width=.5\linewidth]{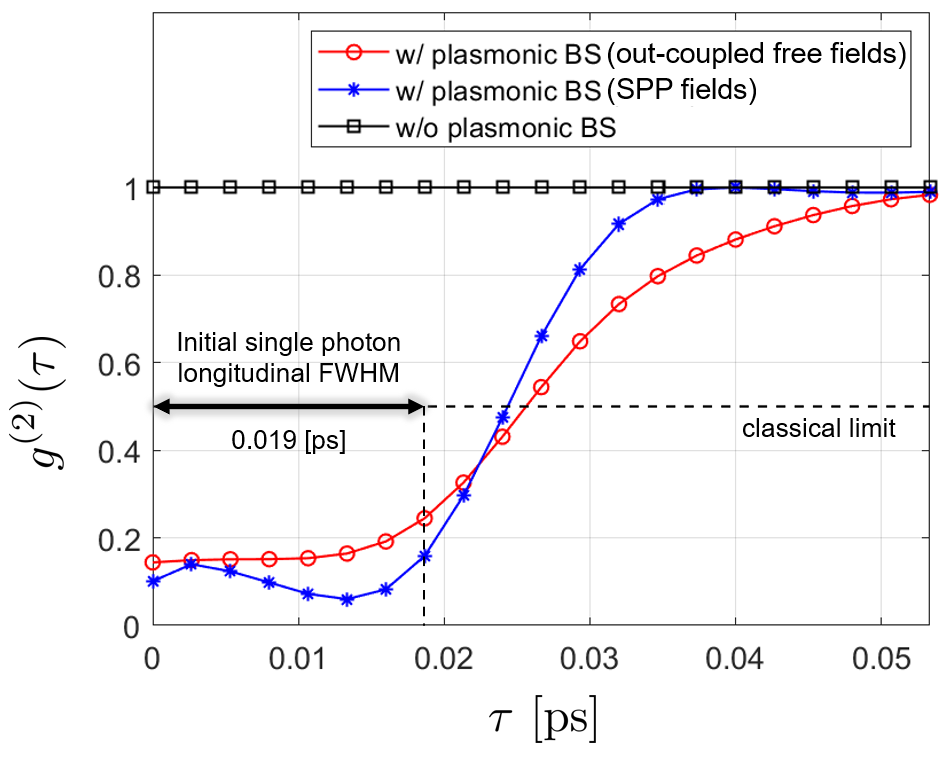}
\caption{Second-order correlation versus delay $\tau$ for three cases:  
(i) plasmonic BS with coincidence of out-coupled fields;  
(ii) plasmonic BS with coincidence of SPP fields;  
(iii) absence of the BS with coincidence of out-coupled fields.}
\label{fig:QP_HOM_result}
\end{figure}

Fig.~\ref{fig:QP_HOM_result} presents the computed second-order correlation $g^{(2)}(\tau)$ as a function of the relative delay $\tau$ between the two input photons for three cases:  
(i) coincidence of out-coupled free fields with the plasmonic BS present,  
(ii) coincidence of SPP fields with the BS present, and  
(iii) coincidence of out-coupled fields in the absence of the BS.  
Cases (i) and (ii) clearly exhibit HOM dips for small~$\tau$ (near the temporal FWHM of the photon wavepacket), whereas case (iii) displays almost unity correlation, as expected.  
The non-zero value at $\tau=0$ arises from Ohmic loss in gold and the limited bandwidth over which the plasmonic BS achieves a true $50{:}50$ splitting ratio.

It is noteworthy that case (ii) exhibits a slightly narrower HOM dip (larger effective quality factor) than case (i).  
This occurs because the out-coupling process introduces additional dispersion and attenuation beyond those present in pure SPP propagation.  
Evaluating second-order correlations from SPP-field coincidences is possible only in numerical simulation,  
as direct measurement of bosonic statistics of SPPs remains extremely challenging experimentally.  
As quantum information science advances and integrated photonic--plasmonic circuits mature,  
fully quantum-theoretic numerical simulations such as ours will play an increasingly important role.  
They provide access to physical observables that are difficult or impossible to measure experimentally,  
similar to recent advances in full-wave macroscopic circuit-QED solvers \cite{9540257,Roth2022Full} 
that have enabled improved design of superconducting quantum hardware.  
Likewise, the present numerical framework enables quantitative analysis of open and dissipative quantum-optics problems involving arbitrary geometries and strongly inhomogeneous media.

\section{Atom--field interaction}

\subsection{Multimode Jaynes--Cummings model combined with the M--LN formalism}

The multimode Jaynes--Cummings (MMJC) model generalizes atom--field interaction by including a continuum of EM field modes.  
It employs the two-level system (TLS) approximation in which the atom's infinite ladder of energy levels is truncated to only the ground and excited states that are resonant with the relevant field modes.  
This approximation relies on the assumption that all other atomic transitions are sufficiently detuned so that their contributions to the dynamics can be neglected.  
By reducing the infinite-dimensional atomic Hilbert space to an effective two-dimensional subspace, the MMJC model provides a computationally efficient framework for describing spontaneous emission and light--matter coupling.

In the weak- and moderately strong-coupling regimes, the system Hamiltonian under the rotating-wave approximation (RWA)~\cite{gerry2023introductory} takes the form
\begin{align}
\hat{H}
&=
\hat{H}_{\text{atom}}
+
\hat{H}_{\text{field}}
+
\hat{H}_{\text{interaction}}
\nonumber \\
&=
\hbar\omega_{a}\hat{\sigma}_{+}\hat{\sigma}_{-}
+
\int_{0}^{\infty} d\omega
\int_{\mathcal{D}_{\lambda}} d\lambda\;
\hbar\omega\,\hat{a}^{\dagger}_{\omega,\lambda}\hat{a}_{\omega,\lambda}
+
\int_{0}^{\infty} d\omega
\int_{\mathcal{D}_{\lambda}} d\lambda\;
\hbar g_{\omega,\lambda}\,\hat{\sigma}_{+}\hat{a}_{\omega,\lambda}
+ \mathrm{H.c.},
\label{eqn:MMJC}
\end{align}
where $\omega_a$ denotes the TLS transition frequency,  
$\hat{\sigma}_{\pm}$ are the raising and lowering operators of the TLS,  
$\hat{a}_{\omega,\lambda}$ annihilates a field quantum of frequency $\omega$ and mode index $\lambda$,  
and $g_{\omega,\lambda}$ is the atom--field coupling coefficient,
\begin{align}
g_{\omega,\lambda}
= -\,i\,\frac{1}{\hbar}\,
\mathbf{d}_a \cdot \mathbf{E}_{\omega,\lambda}(\mathbf{r}_a),
\end{align}
where $\mathbf{d}_a$ is the atomic dipole moment and $\mathbf{E}_{\omega,\lambda}(\mathbf{r}_a)$ is the mode function evaluated at the TLS position.

Since our primary interest lies in spontaneous emission from the TLS, we restrict the Hilbert space to the single-excitation manifold and consider a general quantum state of the form
\begin{flalign}
\ket{\psi(t)}
=
C(t)\ket{e,\{0\}}
+
\int_{0}^{\infty} d\omega
\int_{\mathcal{D}_{\lambda}} d\lambda\;
D_{\omega,\lambda}(t)\ket{g,1_{\omega,\lambda}},
\label{eqn:Wavevector}
\end{flalign}
where $C(t)$ and $D_{\omega,\lambda}(t)$ are the probability amplitudes for the TLS excitation and the single-quantum field excitation, respectively.

The time evolution of this state is governed by the Schr\"{o}dinger equation,
\begin{flalign}
i\hbar\,\frac{d}{dt}\ket{\psi(t)}
=
\hat{H}\ket{\psi(t)}.
\label{eqn:QSE}
\end{flalign}
Substituting \eqref{eqn:MMJC} and \eqref{eqn:Wavevector} into \eqref{eqn:QSE} leads to two coupled equations for the probability amplitudes:
\begin{flalign}
i\frac{d}{dt}C(t)
&=
\omega_a C(t)
+
\int_{0}^{\infty} d\omega
\int_{\mathcal{D}_{\lambda}} d\lambda\;
g_{\omega,\lambda}\,D_{\omega,\lambda}(t),
\label{eqn:C_org}
\\[2mm]
i\frac{d}{dt}D_{\omega,\lambda}(t)
&=
\omega\;D_{\omega,\lambda}(t)
+
g^{*}_{\omega,\lambda}\,C(t).
\label{eqn:D_org}
\end{flalign}
These equations describe the exchange of a single quantum between the TLS and the continuum of field modes.

\subsection{Computationally efficient approach}

Directly solving \eqref{eqn:D_org} requires knowledge of the entire continuum of BA--MA field modes.  
For numerical implementation, these continuum modes must be sampled appropriately, and such sampling becomes increasingly nontrivial in higher-dimensional geometries, in contrast to the tractable one-dimensional case \cite{choi2025non}.  
In \cite{choi2025atom}, a computationally efficient approach was developed by expressing the dynamics in terms of a memory kernel linked to the imaginary part of the dyadic Green's function.  
This formulation enables the dynamics to be computed using standard CEM solvers---such as auxiliary differential equation finite-difference time-domain with a total-field/scattered-field formulation (ADE-FDTD-TF/SF) or the finite-element method (FEM)---without explicitly constructing the BA--MA modal continuum.
Here, we propose an alternative computationally efficient strategy that further accelerates the numerical treatment of the TLS dynamics within the M--LN framework.

Multiplying \eqref{eqn:D_org} by $g_{\omega,\lambda}$ and integrating over $\lambda$ yield
\begin{flalign}
i 
\frac{d}{dt}
\int_{\mathcal{D}_{\lambda}} d\lambda\, 
g_{\omega,\lambda} D_{\omega,\lambda}(t)
&=
\omega 
\int_{\mathcal{D}_{\lambda}} d\lambda\, 
g_{\omega,\lambda} D_{\omega,\lambda}(t)
+
C(t)
\int_{\mathcal{D}_{\lambda}} d\lambda\, 
|g_{\omega,\lambda}|^2.
\label{eqn:D_org2}
\end{flalign}

We now define  
\begin{flalign}
E_{\omega}(t)
\triangleq
\int_{\mathcal{D}_{\lambda}} d\lambda\,
g_{\omega,\lambda} D_{\omega,\lambda}(t),
\label{eqn:E_w_def_1}
\end{flalign}
which allows \eqref{eqn:C_org} and \eqref{eqn:D_org} to be rewritten as
\begin{flalign}
i\frac{d}{dt}C(t)
&=
\omega_a C(t)
+
\int_{0}^{\infty} d\omega\, E_{\omega}(t),
\label{eqn:C_mod}
\\[1mm]
i\frac{d}{dt}E_{\omega}(t)
&=
\omega\,E_{\omega}(t)
+
\Gamma(\omega)\,C(t),
\label{eqn:E_mod}
\end{flalign}
where the coefficient
\begin{flalign}
\Gamma(\omega)
&=
\int_{\mathcal{D}_{\lambda}} d\lambda\, |g_{\omega,\lambda}|^2
=
\frac{1}{\hbar^2}
\int_{\mathcal{D}_{\lambda}} d\lambda\,
|\mathbf{d}_a \cdot \mathbf{E}_{\omega,\lambda}(\mathbf{r}_a)|^2
=
\frac{\hbar\omega^2\mu_0}{\pi}\,
\Big(
\mathbf{d}^{*}
\cdot
\mathrm{Im}\!\left[\overline{\mathbf{G}}_E(\mathbf{r}_a;\mathbf{r}_a,\omega)\right]
\cdot
\mathbf{d}
\Big),
\label{eqn:BA--MA_completeness}
\end{flalign}
is obtained by applying the transverse modal completeness relation of the BA--MA field modes in \eqref{eqn:TMC_BAMA}.

Once the dyadic Green's function is computed over the frequency bandwidth of interest and the frequencies are uniformly sampled with increment $\Delta\omega$, the modified coupled equations \eqref{eqn:C_mod}-\eqref{eqn:E_mod} can be cast into a sparse linear dynamical system:
\begin{flalign}
\frac{d}{dt}\mathbf{X}(t)
=
\overline{\mathbf{M}}
\cdot
\mathbf{X}(t),
\label{eqn:mat_rep_QSE}
\end{flalign}
where
\begin{flalign}
\mathbf{X}(t)
=
\big[
C(t),\,
\sqrt{\Delta\omega}\,E_{\omega_1}(t),\,
\sqrt{\Delta\omega}\,E_{\omega_2}(t),\,
\ldots,\,
\sqrt{\Delta\omega}\,E_{\omega_{N_\omega}}(t)
\big]^{T},
\end{flalign}
and the non-zero entries of the evolution matrix $\overline{\mathbf{M}}$ are
\begin{flalign}
\left[\overline{\mathbf{M}}\right]_{1,1}
&= \omega_a,
\nonumber \\
\left[\overline{\mathbf{M}}\right]_{i,i}
&= \omega_n,
\qquad
n=1,\dots,N_\omega,\;\; i=n+1,
\nonumber \\
\left[\overline{\mathbf{M}}\right]_{1,i}
&= \sqrt{\Delta\omega},
\qquad
i=2,\dots,N_\omega+1,
\nonumber \\
\left[\overline{\mathbf{M}}\right]_{i,1}
&= \Gamma(\omega_n)\sqrt{\Delta\omega},
\qquad
i=2,\dots,N_\omega+1,
\nonumber \\
\left[\overline{\mathbf{M}}\right]_{i,j}
&= 0,\qquad \text{otherwise}.
\end{flalign}

The system \eqref{eqn:mat_rep_QSE} can be integrated numerically using either a fourth-order Runge--Kutta method or the Crank--Nicolson scheme.

It is important to emphasize that the original quantum state is normalized.  
When the continuum variables $(\omega,\lambda)$ are discretized with increments 
$\Delta\omega$ and $\Delta\lambda$, the state vector in the original formulation is
\begin{align}
\mathbf{X}_{\mathrm{orig}}(t)
\triangleq
\big[
C(t),\{\sqrt{\Delta\omega\,\Delta\lambda}\,D_{\omega,\lambda}(t)\}
\big]^{T},
\end{align}
and it must satisfy the normalization condition
\begin{align}
\mathbf{X}_{\mathrm{orig}}^{\dagger}\cdot
\mathbf{X}_{\mathrm{orig}}
\approx 
\left| C(t)\right|^2
+
\int_{0}^{\infty}d\omega \int_{\mathcal{D}_{\lambda}}d\lambda
\left|D_{\omega,\lambda}(t)\right|^2
=
1.
\end{align}
However, the modified representation obeys
\begin{flalign}
\mathbf{X}^{\dagger}(t)\cdot\mathbf{X}(t)
&=
|C(t)|^2
+
\sum_{n=1}^{N_\omega}
|E_{\omega_n}(t)|^2\,\Delta\omega
\nonumber \\
&\approx
|C(t)|^2
+
\int_{0}^{\infty} d\omega\,|E_{\omega}(t)|^2
=
|C(t)|^2
+
\int_{0}^{\infty} d\omega\,
\left|
\int_{\mathcal{D}_{\lambda}} d\lambda\,
g_{\omega,\lambda}D_{\omega,\lambda}(t)
\right|^2
\neq 1.
\end{flalign}

Thus, the evolution matrix $\overline{\mathbf{M}}$ is generally non-unitary (and non-Hermitian).  
This non-unitarity does not reflect physical dissipation but rather arises from the reformulation of the original problem into a reduced computational representation designed to accelerate the numerical algorithm.

\subsection{Single-photon amplitude}
In addition to the TLS dynamics, one may also wish to analyze the out-coupled
field or the radiated field generated by the TLS. 
A natural question then arises: after replacing $D_{\omega,\lambda}(t)$ with the reduced quantity $E_{\omega}(t)$ for computational efficiency, is it still possible to compute the mean electric field or radiation characteristics? Otherwise, such a reduction would render the algorithm ineffective for field-based analyses.

The key point is that the field behavior can indeed be reconstructed from $E_{\omega}(t)$. In fact, the quantity $E_{\omega}(t)$ contains exactly the information required to evaluate the expectation value of the electric field operator at any observation point. The underlying reason is that $E_{\omega}(t)$ is obtained by integrating $g_{\omega,\lambda} D_{\omega,\lambda}(t)$ over the degeneracy index $\lambda$, which is the same
combination that enters the modal expansion of the electric field operator. 
We now explain this in detail below.

Single-photon amplitude is defined by
\begin{flalign}
\mathbf{E}_{\text{spa}}(\mathbf{r},t)
&=
\mel{0}{\hat{\mathbf{E}}(\mathbf{r},0)}{\psi(t)}
=
\int_{0}^{\infty} d\omega
\int_{\mathcal{D}_{\lambda}} d\lambda\;
D_{\omega,\lambda}(t)\,
\mathbf{E}_{\omega,\lambda}(\mathbf{r}),
\label{eq:spa_def}
\end{flalign}
from which the first-order correlation (or single-photon detection probability) in the Schr\"{o}dinger picture follows immediately as
\begin{flalign}
\text{SPDP}(\mathbf{r},t)
=
\mel
{\psi(t)}
{\hat{\mathbf{E}}^{(-)}(\mathbf{r},0)\cdot\hat{\mathbf{E}}^{(+)}(\mathbf{r},0)}
{\psi(t)}
=
\big| \mathbf{E}_{\text{spa}}(\mathbf{r},t) \big|^{2}.
\end{flalign}

Assuming that no photons are initially present, the formal solution of
\eqref{eqn:D_org} is
\begin{flalign}
D_{\omega,\lambda}(t)
=
-i g_{\omega,\lambda}^{*}
\int_{0}^{t} dt'\,
C(t')\,e^{-i\omega(t-t')}.
\label{eqn:D_formal_sol}
\end{flalign}

Substituting \eqref{eqn:D_formal_sol} into \eqref{eq:spa_def}, the single-photon
amplitude can be expressed explicitly in terms of the excited-state amplitude
$C(t)$:
\begin{flalign}
\mathbf{E}_{\text{spa}}(\mathbf{r},t)
&=
-i
\int_{0}^{\infty} d\omega
\int_{\mathcal{D}_{\lambda}} d\lambda\;
\left[
\int_{0}^{t} dt'\,
C(t')e^{-i\omega(t-t')}
\right]
g_{\omega,\lambda}^{*}
\mathbf{E}_{\omega,\lambda}(\mathbf{r}).
\label{eq:spa_C_form_1}
\end{flalign}
This expression shows that the entire emitted field is generated solely by the
temporal evolution of the atomic amplitude \(C(t)\).
Next, using \eqref{eqn:D_formal_sol} in the definition of \(E_\omega(t)\),
we obtain
\begin{flalign}
E_\omega(t)
=
\left[
-i \int_{\mathcal D_\lambda} d\lambda\,
|g_{\omega,\lambda}|^{2}
\right]
\int_{0}^{t} dt'\,
C(t')e^{-i\omega(t-t')}.
\label{eq:Ew_intermediate}
\end{flalign}

The $\lambda$-integration over $|g_{\omega,\lambda}|^{2}$ is evaluated through
the BA--MA transverse modal completeness relation in
\eqref{eqn:BA--MA_completeness}.  
Substituting \eqref{eqn:BA--MA_completeness} into \eqref{eq:Ew_intermediate}
gives the compact expression
\begin{flalign}
E_\omega(t)
=
-i\,\Gamma(\omega)
\int_{0}^{t} dt'\,
C(t')e^{-i\omega(t-t')}.
\label{eq:Ew_final}
\end{flalign}

Finally, inserting \eqref{eq:Ew_final} into \eqref{eq:spa_C_form_1} and applying
the modal completeness relation once more for the spatial part yields the
Green-tensor representation of the emitted single-photon field:
\begin{flalign}
\mathbf{E}_{\text{spa}}(\mathbf{r},t)
=
\frac{\mu_0}{\pi}
\int_{0}^{\infty} d\omega\;
\omega^{2}\,
\Im\!\big[\overline{\mathbf {G}}_E(\mathbf r;\mathbf r_a,\omega)\big]
\cdot \mathbf d_a^{*}\;
\frac{E_\omega(t)}{-i\Gamma(\omega)}.
\label{eq:spa_final}
\end{flalign}





\subsection{Simulation results of spontaneous emission in plasmonic antennas}
\begin{figure}
\centering
\includegraphics[width=.5\linewidth]{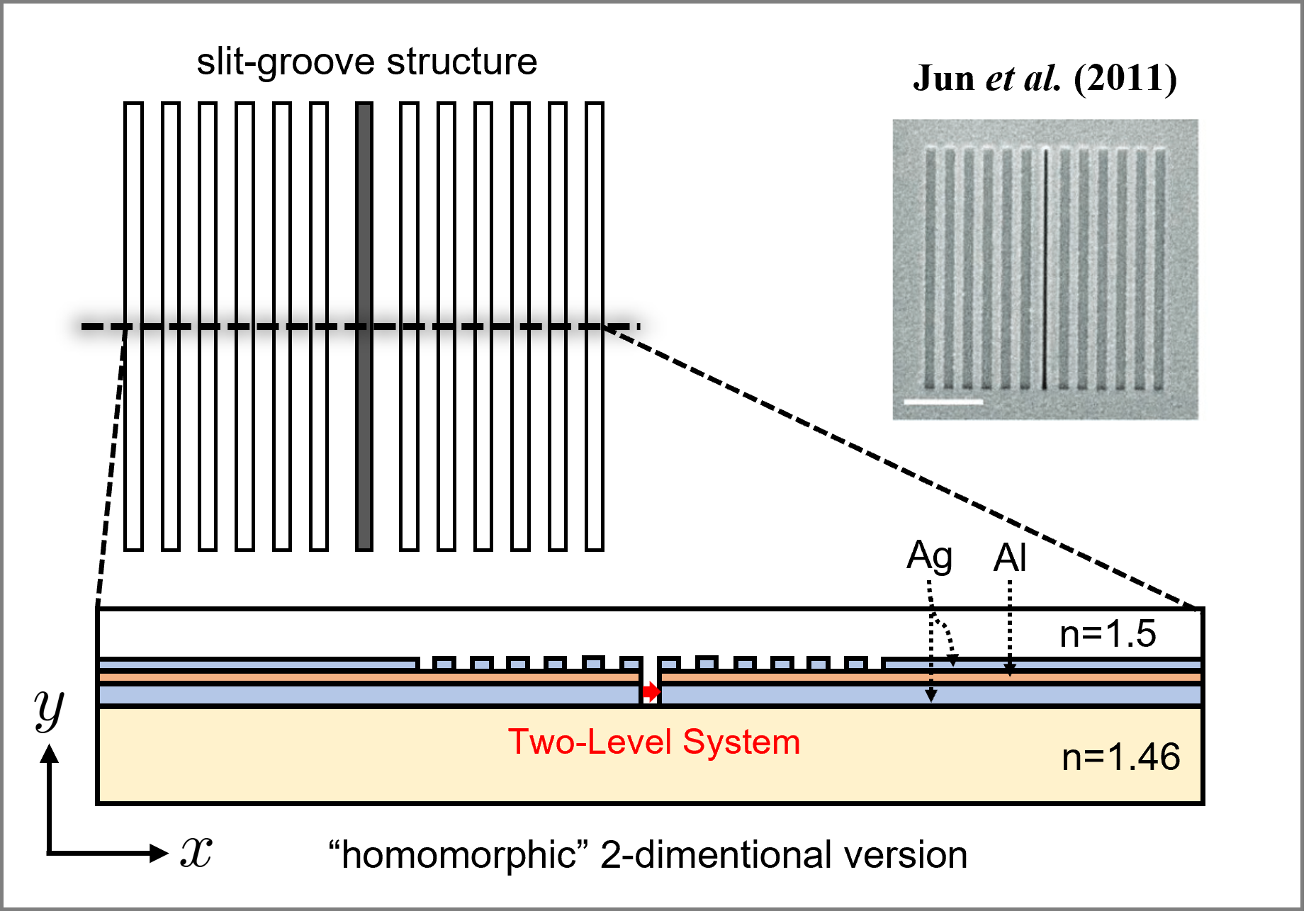}
\caption{A plasmonic slit-groove structure designed to observe the Purcell effect in single-photon emission. The inset figure on the right, illustrating a scanning electron microscope image for the top view of the photon-to-surface-plasmon-polariton launcher, was reproduced from \cite{jun2011plasmonic}. The bottom panel displays the equivalent 2D cross-sectional structure (homomorphic 2D version) used for our simulations, where the red arrow indicates the TLS.}
\label{fig:slit}
\end{figure}

We performed numerical simulations for a TLS placed in the plasmonic slit--groove structures shown in Fig.~\ref{fig:slit}, adopting geometric parameters from \cite{jun2011plasmonic} and material models from \cite{ung2007interference}. 
Our test scenarios include three groove configurations: (1) no grooves, (2) symmetric grooves, and (3) asymmetric grooves.
The simulation geometry consists of a multilayer stack deposited on a dielectric substrate ($n=1.46$). 
The metallic layers comprise Ag (140~nm), Al (40~nm), and Ag (30~nm). 
A central slit with a width of 80~nm is located at the center, flanked by six grooves on each side patterned into the top Ag layer. 
All grooves have a 50\% duty cycle. 
The symmetric structure uses a 310~nm slit--groove distance and 350~nm period, whereas the asymmetric configuration employs periods of 380~nm (left) and 320~nm (right). 
A 340-nm-thick dielectric cladding ($n=1.5$) covers the entire structure. In all three cases, the TLS is positioned within the central slit, 100~nm above the substrate.
To assess the reliability of our numerical framework, we used the spectral function approach (SFA)~\cite{na2023numerical} as a reference. 
The SFA provides the population dynamics predicted solely from the spectral density (i.e., the imaginary part of the dyadic Green's function), which is an exact descriptor of the environmental electromagnetic response. 
Therefore, agreement between our time-domain simulation and the SFA result serves as a stringent validation of our numerical model.
In particular, at early times the TLS dynamics initially follow the spontaneous emission characteristics of free space before the scattered fields return and modify the emitter’s evolution. 
After the system reaches steady state, the TLS population dynamics converge to those predicted by the SFA. 
This transition---from an initial free-space-like decay to an SFA-governed long-time behavior---is a hallmark of non-Markovian atom--field interaction, where the emitter retains memory of its interaction with the structured, dissipative environment.

\begin{figure}
    \centering
    \includegraphics[width=0.5\linewidth]{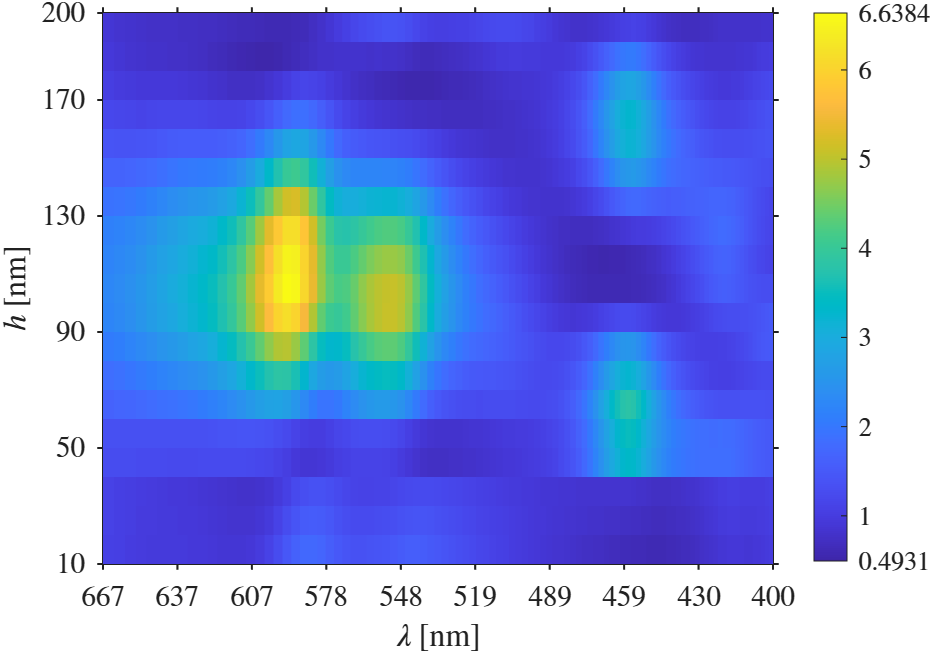}
    \caption{Numerically evaluated Purcell factors for the symmetric slit--groove structure obtained using the SFA. The parameter sweep is performed over the TLS--substrate separation $h$ and the transition wavelength $\lambda$, illustrating how the local EM environment modifies the spontaneous emission rate.}
    \label{fig:purcell_factor_sweep}
\end{figure}

\begin{figure}
\centering
\includegraphics[width=0.5\linewidth]{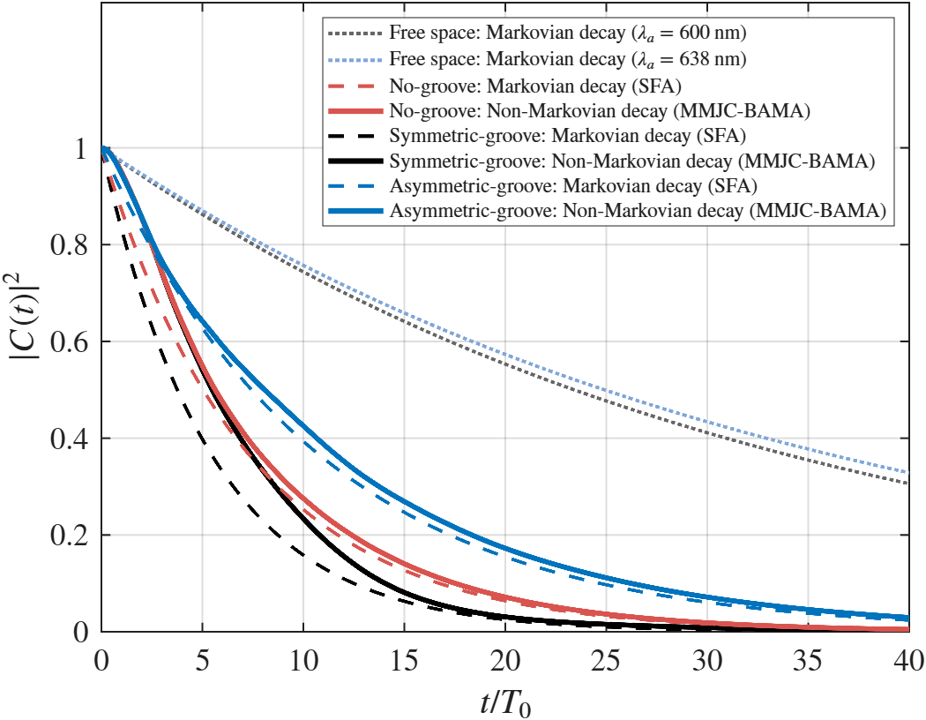}
\caption{Time evolution of the atomic population. 
The results are presented for three cases: (i) slit without grooves; (ii) with symmetric groove structure; and (iii) with asymmetric groove structure. 
The calculations were performed using the SFA and BA-MA methods. 
Simulation parameters: TLS height (distance from the substrate) $h = 100$~nm; atomic transition wavelength $\lambda_a = 600$~nm for (i) and (ii), and $\lambda_a = 638$~nm for (iii). 
Here, $T_0$ denotes the transition period of the TLS.}
\label{fig:population}
\end{figure}

In the SFA, the spontaneous emission rate of a two-level emitter with transition frequency $\omega_a$ and dipole moment $\mathbf{d}=\hat{\mathbf{x}}d$ is determined by the imaginary part of the dyadic Green's function at the emitter's position $\mathbf{r}_a$:
\begin{equation}
\Gamma^{\mathrm{(SFA)}}
= \frac{2\omega_a^2}{\hbar c^2 \epsilon_0}
\mathbf{d} \cdot \mathrm{Im}\left[
\bar{\mathbf{G}}(\mathbf{r}_a; \mathbf{r}_a, \omega_a)
\right] \cdot \mathbf{d}^*
= \frac{2\omega_a^2 d^2 }{\hbar c^2 \epsilon_0}
\mathrm{Im}\left[
G_{x,x}(\mathbf{r}_a, \mathbf{r}_a, \omega_a)
\right].
\end{equation}
In the two-dimensional geometry considered in this work, the Purcell factor is directly related to the local Green's function via
\begin{equation}
F_P = \frac{\Gamma}{\Gamma_0}
= 4\times\mathrm{Im}\left[
G_{x,x}(\mathbf{r}_a; \mathbf{r}_a, \omega_a)
\right].
\label{eq:purcell}
\end{equation}

In order to identify the optimal TLS position that maximizes the Purcell factor, we performed a parametric study with respect to the TLS--substrate separation $h$ and the atomic transition wavelength $\lambda_a$, as shown in Fig.~\ref{fig:purcell_factor_sweep} for the symmetric groove case. From this map, we selected $\lambda_a = 600\,\mathrm{nm}$ and $h = 100\,\mathrm{nm}$, which yield a Purcell factor of approximately 6.6. 
Although the Purcell-factor distribution for the asymmetric groove case is not displayed here, its maximum occurs near $\lambda_a \approx 638\,\mathrm{nm}$ at $h = 100\,\mathrm{nm}$. These parameter sets are used in the subsequent MMJC--BAMA simulations.

Finally, the numerical results obtained from \eqref{eqn:mat_rep_QSE} and \eqref{eq:spa_final} were compared with the atomic population predicted by \eqref{eq:purcell}. 
As shown in Fig.~\ref{fig:population}, the time-evolution analysis captures retardation effects that are absent in the static SFA predictions. 
These deviations arise from the finite propagation time required for the emitted photon to undergo diffraction and reflection before returning to the TLS.

\begin{figure}[t]
\centering

\subfloat[No-groove configuration]{
    \includegraphics[width=0.7\linewidth]{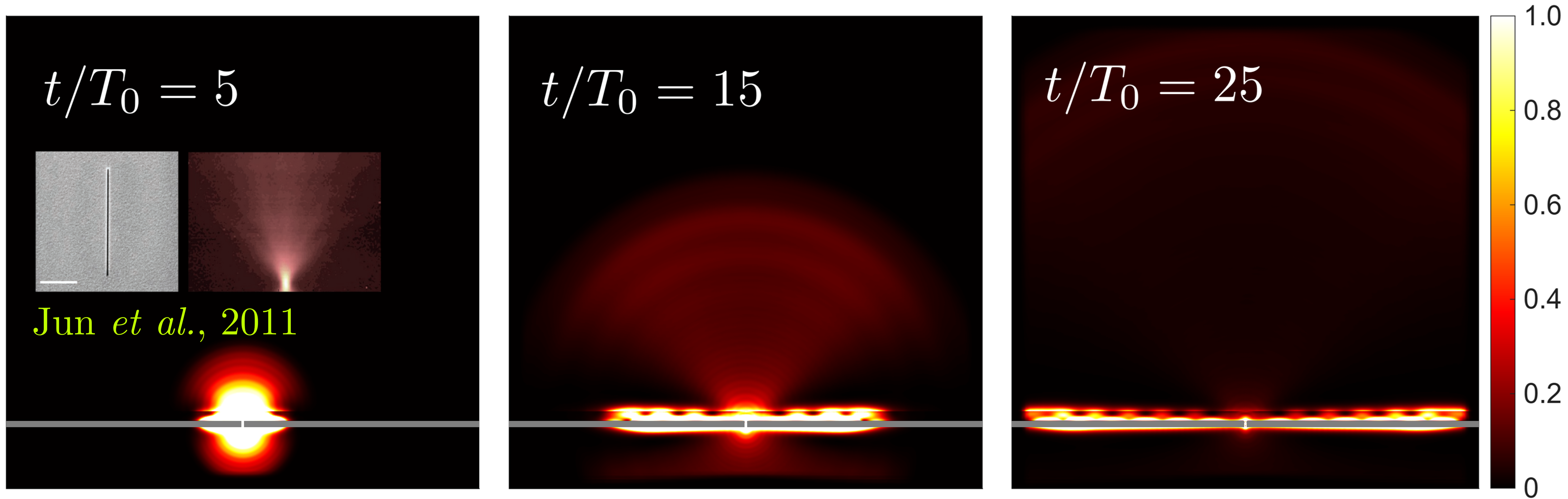}
    \label{fig:time_evo_600nm_nogroove}
}
\\
\subfloat[Symmetric-groove configuration]{
    \includegraphics[width=0.7\linewidth]{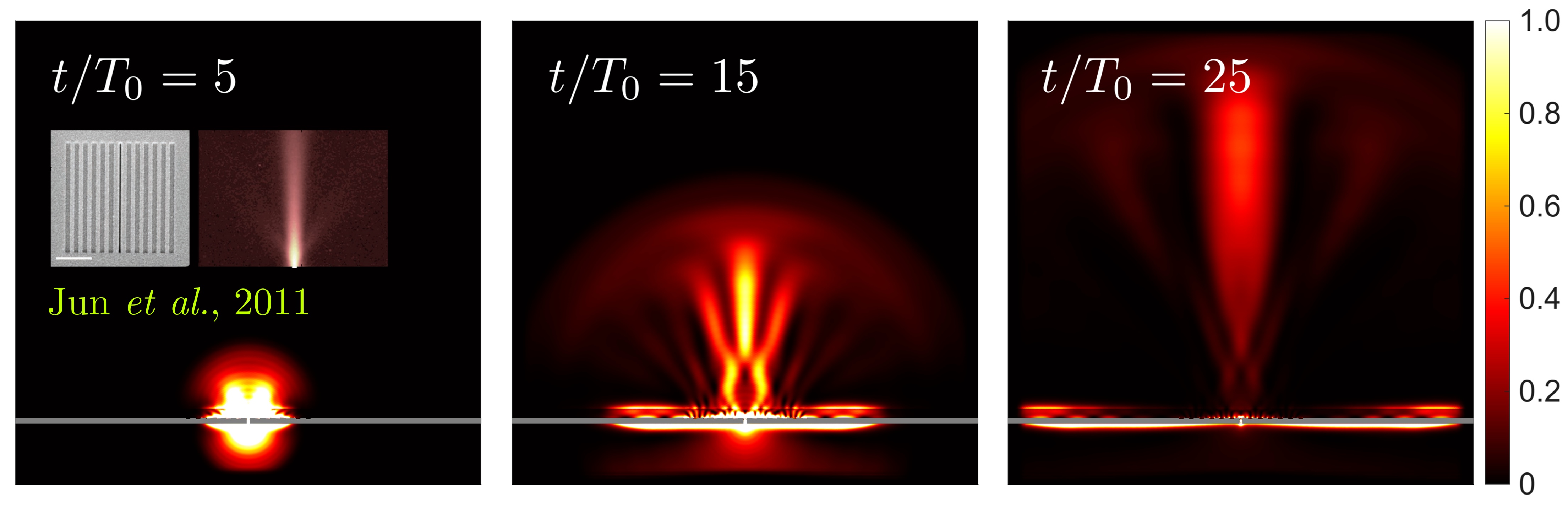}
    \label{fig:time_evo_600nm_symmetric}
}
\\
\subfloat[Asymmetric-groove configuration]{
    \includegraphics[width=0.7\linewidth]{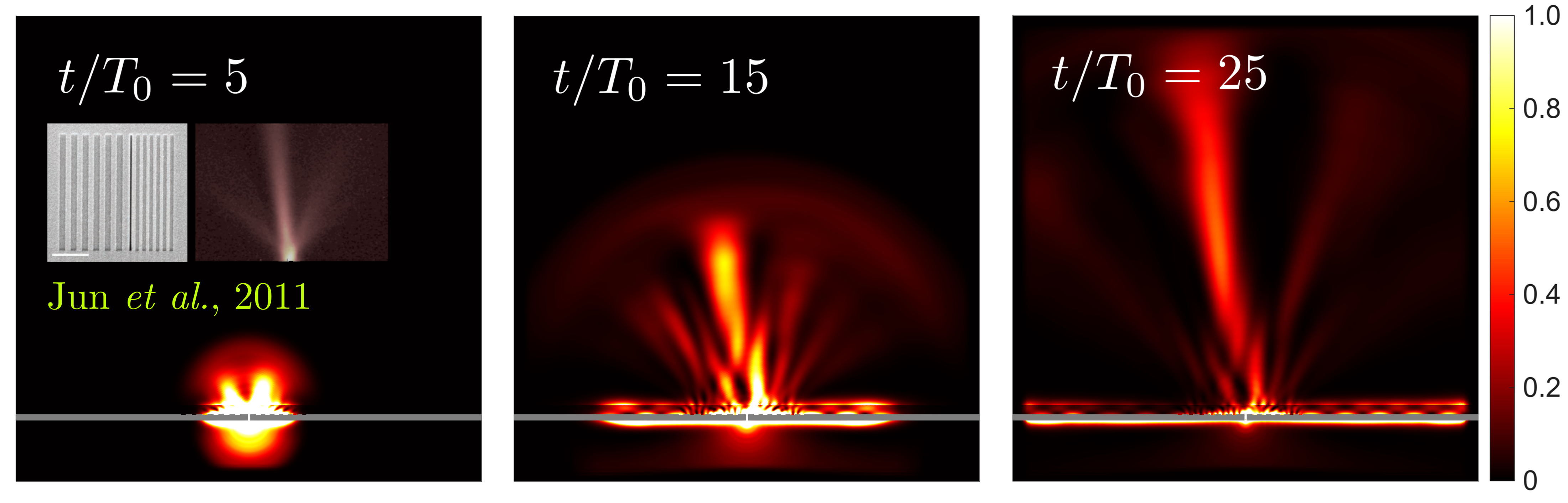}
    \label{fig:time_evo_638nm_asym}
}

\caption{Spatial distributions of the normalized single-photon amplitude 
$|\mathbf{E}_{\mathrm{spa}}|^2$ at $t/T_0 = 5, 15,$ and $25$ for 
(a) the no-groove structure, 
(b) the symmetric slit--groove structure, and 
(c) the asymmetric slit--groove structure. 
The no-groove configuration produces a widely spread single-photon radiation pattern due to diffraction from the subwavelength slit. 
In contrast, the symmetric configuration generates a vertically collimated beam through constructive interference of groove-scattered SPPs, whereas the asymmetric geometry introduces a geometric phase shift that yields a tilted wavefront and directional emission.
}
\label{fig:time_evo_all}
\end{figure}

Fig.~\ref{fig:time_evo_all} displays the normalized single-photon amplitude distributions $|\mathbf{E}_{\mathrm{spa}}|^2$ for the no-groove, symmetric, and asymmetric slit--groove geometries. 
As shown in Fig.~\ref{fig:time_evo_600nm_nogroove}, the no-groove structure radiates into a broad angular distribution, which arises from the diffraction of light emitted through a subwavelength slit---a well-established physical behavior. 
In the symmetric configuration (also in Fig.~\ref{fig:time_evo_600nm_symmetric}), however, constructive interference of groove-scattered SPPs produces a vertically collimated out-coupled beam, effectively focusing the radiation. 
In contrast, the asymmetric configuration in Fig.~\ref{fig:time_evo_638nm_asym} introduces a geometric phase shift among the scattered SPPs, generating a tilted wavefront and directional emission. 
These simulated field patterns agree well with the reference results reported in \cite{jun2011plasmonic}, confirming that the model captures the essential plasmonic interference mechanisms.

\section{Conclusion and Future Works}
In this work, we developed a fully first-principles and computationally efficient framework for modeling quantum optical phenomena in realistic plasmonic environments by leveraging the modified Langevin noise (M--LN) formalism.  
Within the Heisenberg picture, we demonstrated Hong--Ou--Mandel--type two-photon interference in a plasmonic beam-splitting platform by expanding the electric-field operator using BA--MA modes and modeling incident single-photon states through BA-field superpositions.  
A key result is that the second-order correlation and single-photon detection probability can be computed directly from two classical time-domain electromagnetic simulations, providing a practical and scalable route for predicting quantum interference in complex nanostructures.  
Within the Schr\"{o}dinger picture, we proposed a modified multimode Jaynes--Cummings formulation that explicitly incorporates BA--MA coupling while avoiding the computational intractability of sampling the full continuum-mode degeneracy space.  
By lumping out the BA--MA degeneracy degrees of freedom and defining frequency-dependent effective variables, our approach captures atom--plasmon--radiation coupling with high numerical efficiency.  
Using this method, we revealed how plasmonic grating-antenna geometries enable directional and highly structured single-photon emission from a two-level system embedded inside a slit.  
Together, these results establish the M--LN formalism as a versatile, first-principles foundation for quantum plasmonics across both interaction and emission regimes.

The present framework opens several promising avenues for future research.  
First, our method can be extended to multi-emitter systems to investigate cooperative quantum effects arising from atomic inhomogeneous broadening and dipole--dipole interactions mediated by reactive near fields.  
Such an extension would enable the study of superradiance, subradiance, entanglement dynamics, and collective non-Markovian behavior in realistic plasmonic environments involving multiple quantum emitters.
Second, the BA--MA modal quantization offers a natural pathway toward modeling open and dissipative plasmonic systems operating in the ultra-strong and deep-strong coupling regimes.  
Incorporating the minimal-coupling Hamiltonian with its nonlinear terms, combined with tensor-network approaches such as matrix-product-state representations, may enable efficient simulation of strongly interacting light--matter systems beyond the rotating-wave and Markov approximations.  
This direction would provide a rigorous theoretical basis for exploring nonperturbative quantum plasmonic phenomena.
Third, the proposed framework can be generalized to photonic integrated circuit platforms containing plasmonic cavity quantum electrodynamics components.  
Because the BA--MA formalism integrates seamlessly with large-scale computational electromagnetic solvers, it is well suited for scalable modeling of on-chip quantum plasmonic resonators, waveguides, and hybrid plasmonic--dielectric microcavities.  
This would support device-level design and optimization of integrated quantum photonic architectures in realistic open and lossy environments.
Overall, these extensions point toward a broad and impactful research program aimed at understanding and engineering quantum plasmonic systems across multi-emitter, ultra-strong coupling, and chip-scale photonic platforms.

\medskip
\textbf{Supporting Information} \par 
Supporting Information is available from the Wiley Online Library or from the author.

\medskip
\textbf{Acknowledgments} \par 
This work is funded by the Institute for Information \&
Communications Technology Promotion (IITP), the Korea
Government (MSIT) (Grants No. RS-2025-02307012 and No. RS-2025-25464763), and by National Science Foundation Grant No. 2202389.

\medskip

%
\bibliographystyle{MSP}
\bibliography{sample}


\end{document}